\documentclass[12pt,twoside]{article}
\usepackage{amsmath,amsfonts,amssymb}
\usepackage{latexsym}
\usepackage{dcolumn}
\usepackage{graphicx,epsfig}
\usepackage{amsthm}

\begin{document}

\begin{flushright}
{\tt IISER(Kolkata)/GR-QC\\ \today}
\end{flushright}
\vspace{1.5cm}

\begin{center}
{\Large \bf Effects of GUP in Quantum Cosmological Perfect Fluid Models }
\vglue 0.5cm
Barun Majumder\footnote{barunbasanta@iiserkol.ac.in}
\vglue 0.6cm
{\small {\it Department of Physical Sciences,\\Indian Institute of Science Education and Research (Kolkata),\\
Mohanpur, Nadia, West Bengal, Pin 741252,\\India}}
\end{center}
\vspace{.1cm}

\begin{abstract} 
Very recently authors in \cite{b1} proposed a new Generalized Uncertainty Principle (or GUP) with a linear term in Plank length. In this Letter the effect of this GUP is studied in quantum
cosmological models with dust and cosmic string as the perfect fluid. For the quantum mechanical description it is possible to find the wave packet which resulted from the superposition of
the stationary wave functions of the Wheeler-deWitt equation. However the norm of the wave packets turned out to be time dependent and hence the model became non-unitary. The loss of unitarity is
due to the fact that the presence of the linear term in Plank length in the Generalized Uncertainty Principle made the Hamiltonian non-Hermitian. 
\vspace{5mm}\newline Keywords: quantum cosmology, GUP, minimal length
\end{abstract}
\vspace{.3cm}

\section{Introduction}

The idea that the uncertainty principle could be affected by gravity was first given by Mead \cite{c1}. Later modified commutation relations between position and momenta commonly
known as Generalized Uncertainty Principle ( or GUP ) were given by candidate theories of quantum gravity ( String Theory, Doubly Special
Relativity ( or DSR ) Theory and Black Hole Physics ) with the prediction of a minimum measurable length \cite{b2,b7}. Similar kind of
commutation relation can also be found in the context of Polymer Quantization in terms of Polymer Mass Scale \cite{c}.
\par
The authors in \cite{b1} proposed a GUP which is consistent with DSR theory, String theory and Black Hole Physics and which says
\begin{equation}
\left[x_i,x_j\right] = \left[p_i,p_j\right] = 0 ,
\end{equation}
\begin{equation}
\label{g2}
[x_i, p_j] = i \hbar \left[  \delta_{ij} -  l  \left( p \delta_{ij} +
\frac{p_i p_j}{p} \right) + l^2  \left( p^2 \delta_{ij}  + 3 p_{i} p_{j} \right)  \right],
\end{equation}
\begin{align}
\label{g3}
 \Delta x \Delta p &\geq \frac{\hbar}{2} \left[ 1 - 2 l <p> + 4 l^2 <p^2> \right]  \nonumber \\
& \geq \frac{\hbar}{2} \left[ 1  +  \left(\frac{l}{\sqrt{\langle p^2 \rangle}} + 4 l^2  \right)  \Delta p^2  +  4 l^2 \langle p \rangle^2 -  2 l \sqrt{\langle p^2 \rangle} \right],
\end{align}
where $ l=\frac{l_0 l_{pl}}{\hbar} $. Here $ l_{pl} $ is the Plank length ($ \approx 10^{-35} m $). It is normally assumed that the dimensionless
parameter $l_0$ is of the order unity. If this is the case then the $l$ dependent terms are only important at or near the Plank
regime. But here we expect the existence of a new intermediate physical length scale of the order of $l \hbar = l_0 l_{pl}$. We also note
that this unobserved length scale cannot exceed the electroweak length scale \cite{b1} which implies $l_0 \leq 10^{17}$. These equations are
approximately covariant under DSR transformations but not Lorentz covariant \cite{b7}. These equations also imply
\begin{equation}
\Delta x \geq \left(\Delta x \right)_{min} \approx l_0\,l_{pl}
\end{equation}
and
\begin{equation}
\Delta p \leq \left(\Delta p \right)_{max} \approx \frac{M_{pl}c}{l_0}
\end{equation}
where $ M_{pl} $ is the Plank mass and $c$ is the velocity of light in vacuum. It can be shown that equation (\ref{g2}) is satisfied by the
following definitions $x_i=x_{oi}$ and $p_i=p_{oi} (1 - l\,p_o + 2\,l^2\,p_o^2)$, where $x_{oi}$, $p_{oj}$ satisfies $[x_{oi}, p_{oj}]= i \hbar \delta_{ij}$. Here we can interpret $p_{oi}$ as the momentum at low energies having the standard representation in position space ($ p_{oi}=-i\hbar \frac{\partial}{\partial x_{oi}}$) with $p_o^2=\sum_{i=1}^3 p_{oi}p_{oi}$ and $ p_i $ as the momentum at high energies. We can also show that the $ p^2 $ term in the kinetic part of any Hamiltonian can be written as \cite{b1}
\begin{equation}
\label{corr}
 p^2 \Longrightarrow \ p_o^2 - 2\ l\ p_o^3 + {\cal O}(l^2) + \ldots \, \, \, .
\end{equation}
Here we neglect terms $ {\cal O}(l^2)$ and higher in comparison to terms $ {\cal O}(l)$ to study the effect of the linear term in $l$ in the first approximation as $l=l_0\,l_{pl}$. The effect of this proposed GUP is studied for some well known quantum
mechanical Hamiltonians in \cite{b1,b8}. 
\par 
In this Letter we are going to study the effect of this GUP \cite{b1} ( only upto a linear term in $l$ ) in some selected quantum cosmological perfect fluid models with dust and
cosmic string. For brief discussion on quantum cosmological perfect fluid models we can see \cite{alv,d10,d12,d11,d14,f,my}.

\section{Quantum Perfect Fluid Cosmological Models}

The expression for action in these quantum cosmological models with perfect fluid can be written as
\begin{equation}
{\cal A} = \int_Md^4x\sqrt{-g}\ R + 2\int_{\partial M}d^3x\sqrt{h}\ h_{ab}\ K^{ab} + \int_Md^4x\sqrt{-g}\ P ,
\end{equation}
where $h_{ab}$ is the induced metric over three dimensional spatial hypersurface which is the boundary $\partial M$ of the four dimensional manifold M and $K^{ab}$ is the extrinsic
curvature. Here units are so chosen that $c=16\pi G=\hbar=1$. The second term was first obtained in \cite{d2}. $P$ is the pressure of the fluid and satisfies
the equation of state $P=\alpha \rho$ where $\rho$ is the energy density and $-1\leq\alpha<1$. In Schutz's formalism \cite{d8,d9} the fluid's four velocity can be expressed
in terms of three potentials $\epsilon$, $\theta$ and $S$ (here we are studying spatially flat FRW model so other potentials are absent in this model because of its symmetry),
\begin{equation}
u_\nu = \frac{1}{h}(\epsilon_{,\nu} + \theta S_{,\nu}).
\end{equation}
Here $h$ is the specific enthalpy, $S$ is the specific entropy, $\epsilon$ and $\theta$ have no direct physical
meaning. The four velocity also satisfy the normalization condition
\begin{equation}
u^\nu u_\nu = 1.
\end{equation}
The metric for the spatially flat FRW model is
\begin{equation}
ds^2=N^2(t)dt^2-a^2(t)\left[{dr^2}+r^2(d\vartheta^2+ \sin^2 \vartheta d\varphi^2)\right],
\end{equation}
where $N(t)$ is the lapse function and $a(t)$ the scale factor.
Using Schutz's formalism \cite{d8,d9} along with some thermodynamic considerations \cite{d11} it is possible to simplify the action. The final form of the super-Hamiltonian after
using some canonical transformations \cite{d11,alv} can be written as
\begin{equation}
\label{sh}
{\cal H} = N\left[-\frac{p_a^2}{24a} + \frac{p_T}{a^{3\alpha}}\right].
\end{equation}
The lapse function $N$ plays the role of a Lagrange multiplier leading to the constraint ${\cal H}=0$. Here the only canonical variable associated with
matter is $p_T$ and it appears linearly in the super-Hamiltonian. The equation of motion $\dot{T} = \frac{\partial {\cal H}}{\partial p_T} = N a^{-3\alpha} $ reveals that in
the gauge $ N = a^{3\alpha}$, $T$ may play the role of cosmic time. Using usual quantization procedure we can get the Wheeler-deWitt equation for our super-Hamiltonian believing
that the super-Hamiltonian operator annihilates the wave function. So with $ p_a \rightarrow -i\partial_a$ , $ p_T \rightarrow i\partial_t$ and $ \hat{{\cal H}} \Psi (a,t) = 0$ we get
\begin{equation}
\label{weq}
\frac{\partial^2 \Psi}{\partial a^2} + i 24 a^{(1-3\alpha)}\frac{\partial \Psi}{\partial t} = 0.
\end{equation}
Here we have considered a particular choice of factor ordering and our final results will be independent of the different choices of factor ordering. Any two
wave functions $\Phi$ and $\Psi$ must take the form \cite{d14,d12,alv}
\begin{equation}
\label{norm}
\langle \Phi | \Psi \rangle = {\int_0}^\infty a^{(1-3\alpha)} \Phi^* \Psi da
\end{equation}
to make the Hamiltonian operator self-adjoint and the restrictive boundary conditions being 
\begin{equation}
\label{bc}
\Psi(0,t) = 0 \quad \mbox{or} \quad \left. \frac{\partial \Psi(a,t)}{\partial a}\right|_{a=0} = 0.
\end{equation}
To solve equation (\ref{weq}) we can use the method of separation of variables. Writing 
\begin{equation}
\Psi (a,t) = e^{-iEt}\phi (a)
\end{equation}
and using (\ref{weq}) we get
\begin{equation}
\frac{\partial^2 \phi}{\partial a^2} + 24 E a^{(1-3\alpha)} \phi = 0.
\end{equation}
The solutions of this equation can be written in terms of Bessel functions and we can now write the stationary wave functions as
\begin{equation}
\Psi_E = e^{-iEt}\sqrt{a}\biggr[c_1J_{\frac{1}{3(1 - \alpha)}}\biggr(\frac{\sqrt{96E}}{3(1 - \alpha)}a^{\frac{3(1 - \alpha)}{2}}\biggl) + 
c_2Y_{\frac{1}{3(1 - \alpha)}}\biggr(\frac{\sqrt{96E}}{3(1 - \alpha)}a^{\frac{3(1 - \alpha)}{2}}\biggl)\biggl] \quad 
\end{equation}
where $c_{1,2}$ are the integration constants. To satisfy the first boundary condition of (\ref{bc}) we consider $c_1\neq 0$ and $c_2=0$ (to avoid the divergence of
the wave function in the limit $a\rightarrow0$), but still these solutions do not have finite norm. So we are interested in constructing the wave packet by superposing these
solutions. In doing so we consider that the integration constant $c_1$ to be a gaussian function of the parameter $E$. Setting $s=\frac{\sqrt {96E}}{3(1-\alpha)}$ the expression for the
wave packet can be written as 
\begin{align}
\Psi(a,t) &= \sqrt{a}\int_0^\infty s^{\nu + 1}\ e^{-\gamma s^2 - i\frac{3}{32}s^{2}
(1 - \alpha)^2t}\ J_\nu(s a^\frac{3(1 - \alpha)}{2})\ ds \nonumber \\
 &=\frac{a}{{(2 \eta)}^{\frac{4-3\alpha}{3(1-\alpha)}}} e^{-\frac{a^{3(1-\alpha)}}{4 \eta}}
\end{align}
where $\eta = \gamma + i\frac{3}{32}(1-\alpha)^2 t$, $\nu = \frac{1}{3(1-\alpha)}$ and $\gamma$ is an arbitrary positive constant in the gaussian factor. To find the
norm of the wave function for $\alpha=0$ (dust) we use equation (\ref{norm}) and we finally get
\begin{align}
\label{e18}
\langle \Psi | \Psi \rangle &= {\int_0}^\infty a \Psi^* \Psi da \nonumber \\
&=\frac{\Gamma(\frac{4}{3})}{3 (2\gamma)^{\frac{4}{3}}} \quad .
\end{align}
So the norm is finite and independent of time. Similarly for $\alpha=-\frac{1}{3}$(cosmic string) we see that
\begin{align}
\label{e19}
\langle \Psi | \Psi \rangle &= {\int_0}^\infty a^2 \Psi^* \Psi da \nonumber \\
&=\frac{\Gamma(\frac{5}{4})}{4 (2\gamma)^{\frac{5}{4}}} \quad 
\end{align}
which is also finite and time independent.

\section{Effect of the Generalized Uncertainty Principle in these Quantum Cosmological models}

Now we are going to study the effect of the Generalized Uncertainty Principle (or GUP) in the context of the quantum cosmological models described above. Here we
will study two cases, model with dust as the Schutz's perfect fluid and the model with an equation of state $P=-\frac{\rho}{3}$ (cosmic string). Throughout this whole
process we will keep in mind that equation (\ref{g2}) and (\ref{g3}) have a linear term in Plank length as $l =l_0\, l_{pl}$. So we will neglect terms ${\cal O}(l^2)$ and higher in
the first approximation whenever they appear in the calculation.
Due to GUP the $p_a^2$ term of the super-Hamiltonian (\ref{sh}) should be corrected. Following the arguments in \cite{b1} and using (\ref{corr}) we rewrite (\ref{sh}) as 
\begin{equation}
\label{e20}
{\cal H} = N\left[-\frac{1}{24a} (p_o^2 - 2 l p_o^3) + \frac{p_T}{a^{3\alpha}}\right].
\end{equation}
Here we have neglected terms ${\cal O}(l^2)$. Using usual quantization procedures we find
\begin{equation}
\label{e21}
\frac{\partial^2 \Psi}{\partial a^2} + i 2 l \frac{\partial^3 \Psi}{\partial a^3} + i 24 a^{(1-3\alpha)} \frac{\partial \Psi}{\partial t} = 0.
\end{equation}
Using $\Psi (a,t)= e^{-iEt} \phi(a)$ we separate the variables and we get
\begin{equation}
\label{dust1}
\frac{\partial^2 \phi}{\partial a^2} + i 2 l \frac{\partial^3 \phi}{\partial a^3} + 24 E a^{(1-3\alpha)}\phi = 0.
\end{equation}
As mentioned before we will study two cases. One with $\alpha=0$ and another with $\alpha=-\frac{1}{3}$.

\subsection{\underline{ $\alpha = 0$ (Dust) }}

With $\alpha=0$ equation (\ref{dust1}) reduces to 
\begin{equation}
\label{dust2}
\frac{\partial^2 \phi}{\partial a^2} + i 2 l \frac{\partial^3 \phi}{\partial a^3} + 24 E a \phi = 0.
\end{equation}
This third order equation is very difficult to solve analytically. So we will try to solve this equation approximately \cite{bv} in the region $a\approx 0$ (early universe). The solution of
equation (\ref{dust2}) without the $l$ term can be written as
\begin{equation}
\label{sd1}
\phi = d_1 \sqrt{a}\, J_{\frac{1}{3}}\bigg(\sqrt{\frac{32E}{3}} a^{\frac{3}{2}}\bigg) ,
\end{equation}
where $d_1$ is one integration constant while there is a second one which is assigned to the Bessel function of second kind, i.e. $Y_{\frac{1}{3}}$, and is set to zero to avoid the divergence in small $a$ limit. As we are studying early universe cosmology so in
the region $a\approx 0$ (\ref{sd1}) can be written as \cite{bell}
\begin{align}
\phi &\approx d_1 \sqrt{a}\Bigg[\frac{1}{\Gamma\big(\frac{4}{3}\big)}\bigg(\sqrt{\frac{8E}{3}}\bigg)^{\frac{1}{3}}a^{\frac{1}{2}} - 
\frac{1}{\Gamma \big(\frac{7}{3}\big)}\bigg(\sqrt{\frac{8E}{3}}\bigg)^{\frac{7}{3}}a^{\frac{7}{2}} + \ldots \Bigg] \nonumber \\
&\approx D_1 a - D_2 a^4 \,,  
\label{d25}
\end{align}
where $D_1 =d_1 \frac{1}{\Gamma\big(\frac{4}{3}\big)}\bigg(\sqrt{\frac{8E}{3}}\bigg)^{\frac{1}{3}}$ and
 $D_2 = d_1 \frac{1}{\Gamma \big(\frac{7}{3}\big)}\bigg(\sqrt{\frac{8E}{3}}\bigg)^{\frac{7}{3}}$. So clearly $\frac{\partial^3 \phi}{\partial a^3} = -24 D_2 a$. From (\ref{d25}) we
see that for small $a$ we can also consider the approximation $\phi \approx D_1 a$ and the result we get is 
\begin{equation}
\frac{\partial^3 \phi}{\partial a^3} = -48 E \phi \, .
\end{equation} 
If we incorporate this result in equation (\ref{dust2}) we get 
\begin{equation}
\frac{\partial^2 \phi}{\partial a^2} + 24 E a \phi - i 96 l E \phi = 0.
\end{equation}
The solution of this equation is known in terms of Bessel functions and we can write the final form of the stationary wave functions as
\begin{equation}
\Psi_E = c_1 e^{-iEt} \, \sqrt{a-i4l}\,J_{\frac{1}{3}}\bigg(\frac{2}{3}\sqrt{24E}\,(a-i4l)^{\frac{3}{2}}\bigg)
\end{equation}
where $c_1$ is one integration constant while there is a second one which is assigned to the Bessel function of second kind, i.e. $Y_{\frac{1}{3}}$, and is set to zero to avoid the divergence in small $a$ limit. In this case also we should construct the wave packet superposing these solutions.
So for the wave packet we can write
\begin{equation}
\label{e29}
\Psi (a,t) = \int_0^\infty A(E) \Psi_E(a,t)\, dE.
\end{equation}
Defining $s=\frac{2}{3}\sqrt{24E}$ and considering $A(E)$ to be a gaussian function (here we have chosen $A = \frac{16}{3} s^{\frac{1}{3}} e^{-\gamma s^2}$), the expression for the wave packet can be written as
\begin{equation}
 \Psi (a,t) = \sqrt{a-i4l}\,\int_0^\infty e^{-s^2\big(\gamma+i\frac{3}{32}t\big)}\,s^{\frac{4}{3}}\,J_{\frac{1}{3}}\big(s\,(a-i4l)^{\frac{3}{2}}\big)\,ds \,.
\end{equation}
This is a known integral \cite{bell} and finally we can write
\begin{equation}
\Psi(a,t) = \frac{(a-i4l)}{2^{\frac{4}{3}}\big(\gamma+i\frac{3}{32}t\big)^{\frac{4}{3}}} \, e^{-\frac{(a-i4l)^3}{4\big(\gamma+i\frac{3}{32}t\big)}}\,.
\end{equation}
A straightforward calculation gives
\begin{equation}
\label{e32}
\Psi^*\,\Psi = (2A)^{-\frac{8}{3}}\,a^2\,e^{-\frac{\gamma}{2A^2}a^3}\,e^{\frac{9lt}{16A^2}a^2}
\end{equation}
where $A=\big(\gamma^2+\frac{9}{1024}t^2\big)^{\frac{1}{2}}$. As we are interested in the norm of the wave packet we have to follow equation (\ref{norm}) and in this case we have to evaluate
\begin{equation}
\langle \Psi | \Psi \rangle = \int_0^\infty a\, \Psi^*\,\Psi\, da.
\end{equation}
Using equation (\ref{e32}) we evaluate the square of the norm as 
\begin{align}
 \label{e34}
\langle \Psi | \Psi \rangle &= (2A)^{-\frac{8}{3}}\,\int_0^\infty\,a^3\,e^{-\frac{\gamma}{2A^2}a^3}\,e^{\frac{9lt}{16A^2}a^2}\,da \nonumber \\
&= \frac{\Gamma(\frac{4}{3})}{3(2\gamma)^{\frac{4}{3}}} + \frac{3}{32}2^{\frac{1}{3}} \, \frac{lt}{\gamma^2\big(\gamma^2+\frac{9}{1024}t^2\big)^{\frac{1}{3}}}\,.
\end{align}
Throughout this whole process we have neglected all the terms ${\cal O}(l^2)$ and higher. Clearly we can see from equation (\ref{e34}) that the norm is time dependent and hence we can conclude
that this quantum model is non-unitary. If we set $l=0$ we will get back equation (\ref{e18}) and there the norm is time independent. So, keeping in mind this interesting result let us study
the quantum model with cosmic string as the perfect fluid.

\subsection{\underline{ $\alpha = -\frac{1}{3}$ (Cosmic String) }}

If we consider a cosmic string fluid then equation (\ref{dust1}) reduces to
\begin{equation}
\label{e35}
 \frac{\partial^2 \phi}{\partial a^2} + i 2 l \frac{\partial^3 \phi}{\partial a^3} + 24 E a^2 \phi = 0.
\end{equation}
Approaching in the same way as we did in the dust case we can write 
\begin{equation}
\label{e36}
 \phi = d_1\, \sqrt{a}\,J_{\frac{1}{4}}\big(\sqrt{6E}\,a^2\big)
\end{equation}
for $l=0$. In the limit $a\rightarrow0$ equation (\ref{e36}) can be expanded as 
\begin{align}
 \phi &\approx d_1 \frac{1}{\Gamma\big(\frac{5}{4}\big)}\bigg(\frac{3E}{2}\bigg)^{\frac{1}{8}}\,a \,-\, 
d_1\frac{1}{\Gamma \big(\frac{9}{4}\big)}\bigg(\frac{3E}{2}\bigg)^{\frac{9}{8}}\,a^5 + \ldots  \nonumber \\
&\approx D_1 a - D_2 a^5 \,,  
\label{e37}
\end{align}
where $D_1$ and $D_2$ are the coefficients of $a$ and $a^5$ respectively. So clearly $\frac{\partial^3 \phi}{\partial a^3}=-60D_2a^2$. For small enough $a$ the
approximation $\phi\approx D_1a$ yields
\begin{equation}
 \frac{\partial^3 \phi}{\partial a^3}= -72 E a \phi\,.
\end{equation}
Putting this in equation (\ref{e35}) we get
\begin{equation}
\label{e39}
 \frac{\partial^2 \phi}{\partial a^2} + (24Ea^2-i144\,lEa)\,\phi=0\,.
\end{equation}
If we take $x=a-i3\,l$ the equation (\ref{e39}) reduces to 
\begin{equation}
\label{e40}
 \frac{\partial^2 \phi}{\partial x^2} + (24Ex^2+216\,l^2E)\,\phi=0\,.
\end{equation}
Here also we will neglect the term ${\cal O}(l^2)$ and find the solution of equation (\ref{e40}). The solution is known and we now write the final form of the stationary wave functions:
\begin{equation}
 \Psi_E = c_1\, e^{-iEt}\, \sqrt{(a-i3l)}\, J_{\frac{1}{4}}\big(\sqrt{6E}\,(a-i3l)^2\big)\, .
\end{equation}
To construct the wave packet superposing these solutions we have to evaluate equation (\ref{e29}) again in this case. Here we define $s=\sqrt{6E}$ and choose $A(E)$ in such a manner
so that we can
easily do the integration as before. After a straightforward calculation we now write the final form of the wave packet:
\begin{equation}
 \Psi (a,t) = \frac{(a-i3l)}{2^{\frac{5}{4}}\,\big(\gamma+\frac{i}{6}t\big)^{\frac{5}{4}}} \, e^{-\frac{(a-i3l)^4}{4(\gamma+\frac{i}{6}t)}}\,.
\end{equation}
This equation implies
\begin{equation}
 \Psi^*\,\Psi = (2A)^{-\frac{5}{2}}\,a^2\, e^{-\frac{\gamma}{2A^2}a^4}\,e^{\frac{l\,t}{A^2}a^3}
\end{equation}
where $A=\big(\gamma^2+\frac{t^2}{36}\big)^{\frac{1}{2}}$. Now using equation (\ref{norm}) in this case we evaluate the square of the norm of the wave packet and it turns out to be
\begin{align}
 \langle \Psi | \Psi \rangle &= (2A)^{-\frac{5}{2}}\,\int_0^\infty a^4\, e^{-\frac{\gamma}{2A^2}a^4}\,e^{\frac{l\,t}{A^2}a^3} \, da \nonumber \\
&= \frac{\Gamma(\frac{5}{4})}{4(2\gamma)^{\frac{5}{4}}} +  \frac{lt}{2^{\frac{5}{2}}\gamma^2\big(\gamma^2+\frac{t^2}{36}\big)^{\frac{1}{2}}}\,.
\end{align}
In the whole process of the calculation we have neglected terms ${\cal O}(l^2)$ and higher. If $l=0$ we get back equation (\ref{e19}). So this model like the dust model is also non-unitary
as the square of the norm is time dependent.
\par
Anisotropic quantum cosmological models are not unitary as the Hamiltonian operator in those anisotropic models is Hermitian but not self-adjoint \cite{d14,f11,f}. But in our case
if we carefully study equations (\ref{e20}) and (\ref{e21}) we can understand that the effective-Hamiltonian operator which is defined by $H_{eff}=N\big(\frac{\partial^2}{\partial a^2} + i\,2\,l\, 
\frac{\partial^3}{\partial a^3}\big)$ is not Hermitian or very weakly Hermitian in the limit $l\rightarrow0$. So the loss of unitarity is due to the fact that the presence of a linear
term in Plank length in the Generalized Uncertainty Principle is making the effective-Hamiltonian operator non-Hermitian.

\section{Conclusions}

With the very recently proposed Generalized Uncertainty Principle (or GUP) \cite{b1} we have studied the flat minisuperspace FRW quantum cosmological model with dust and cosmic string
as the perfect fluid. This GUP has a linear term in Plank length and here we have studied the effect of this term in the context of very early universe. In both the
cases (dust and cosmic string) Schutz's mechanism has allowed us to obtain the Wheeler-deWitt equation for this minisuperspace in our early universe. Well behaved wave packet can be
constructed from the linear superposition of the stationary wave functions of the Wheeler-deWitt equation. While solving the Wheeler-deWitt equation we considered a particular
choice of factor ordering of the position and momentum operators present in the equation and it is seen that the behaviour of the constructed wave packet remains same for other
factor orderings. The presence of the linear term in Plank length in the GUP made the norm of the wave packet time dependent. So the model became non-unitary. But in the
limit $l_{pl}\rightarrow 0$ the norm becomes time independent.
\par
A standard axiom of quantum mechanics requires that the Hamiltonian should be Hermitian because Hermiticity guarantees that the energy spectrum is real and that time evolution is
unitary (probability-preserving). But here we have seen that the presence of the linear term in Plank length made the Hamiltonian non-Hermitian and as a result total probability is not
conserved. 

\section*{Acknowledgements}
The author is very much thankful to Prof. Narayan Banerjee for helpful discussions and guidance.


\begin{thebibliography}{100}

\bibitem{c1} C. A. Mead, Phys.\ Rev.\  D {\bf 135} (1964) 849.
\bibitem{b2} D. Amati, M. Ciafaloni, G. Veneziano, Phys. Lett. B {\bf 216} (1989) 41;\\
M. Maggiore, Phys.\ Lett.\  B {\bf 304} (1993) 65;\\
M. Maggiore, Phys.\ Rev.\  D {\bf 49} (1994) 5182;\\
M. Maggiore, Phys.\ Lett.\  B {\bf 319} (1993) 83;\\
L. J. Garay, Int.\ J.\ Mod.\ Phys.\  A {\bf 10} (1995) 145;\\
F. Scardigli, Phys.\ Lett.\  B {\bf 452} (1999) 39;\\
S. Hossenfelder, M. Bleicher, S. Hofmann, J. Ruppert, S. Scherer and H. Stoecker, Phys.\ Lett.\  B {\bf 575} (2003) 85;\\
C. Bambi and F. R. Urban, Class.\ Quant.\ Grav.\  {\bf 25} (2008) 095006;\\
A. Kempf, G. Mangano, R. B. Mann, Phys. Rev. D {\bf 52} (1995) 1108;\\
A. Kempf, J.Phys. A  {\bf 30} (1997) 2093;\\
F. Brau, J. Phys. A {\bf  32} (1999) 7691;\\
J. Magueijo and L. Smolin, Phys.\ Rev.\ Lett.\  {\bf 88} (2002) 190403;\\
J. Magueijo and L. Smolin, Phys.\ Rev.\  D {\bf 71} (2005) 026010.
\bibitem{b7} J. L. Cortes, J. Gamboa, Phys. Rev. D {\bf 71} (2005) 065015.
\bibitem{c} G. M. Hossain, V. Husain, S. S. Seahra, Class. Quantum Grav. {\bf 27} (2010) 165013.
\bibitem{b1} A. F. Ali, S. Das and E. C. Vagenas, Phys.\ Lett.\  B {\bf 678} (2009) 497.
\bibitem{b8} S. Das, E. C. Vagenas, Phys. Rev. Lett. {\bf 101} (2008) 221301;\\
 S. Das, E. C. Vagenas, Can. J. Phys. {\bf 87} (2009) 233;\\
 S. Das, E. C. Vagenas, A. F. Ali, Phys.\ Lett.\  B {\bf 690} (2010) 407;\\
 P. Alberto, S. Das, E. C. Vagenas, Phys.\ Lett.\  A {\bf 375} (2011) 1436.
\bibitem{alv} F.G. Alvarenga, J.C. Fabris, N.A. Lemos and G.A. Monerat, Gen. Rel. Grav. {\bf 34} (2002) 651.
\bibitem{d10} M. J. Gotay and J. Demaret, Phys. Rev. D {\bf 28} (1983) 2402;\\
J. Acacio de Barros, N. Pinto-Neto and M. A. Sagioro-Leal, Phys. Let. A {\bf 241} (1998) 229;\\
B. Vakili, Phys. Lett. B {\bf 688} (2010) 129;\\
P. Pedram and S. Jalalzadeh, Phys.Lett. B {\bf 659} (2008) 6.
\bibitem{d12} F. G. Alvarenga and N. A. Lemos, Gen. Rel. Grav.{\bf 30} (1998) 681.
\bibitem{d11} V. G. Lapchinskii and V. A. Rubakov, Theor. Math. Phys. {\bf 33} (1977) 1076.
\bibitem{d14} N. A. Lemos, J. Math. Phys. {\bf 37} (1996) 1449.
\bibitem{f} F.G. Alvarenga, A.B. Batista, J.C. Fabris and S.V.B. Gonsalves, Gen.Rel.Grav. {\bf 35} (2003) 1659.
\bibitem{my} B. Majumder, Phys. Lett. B {\bf 697} (2011) 101.
\bibitem{d2} R. Arnowitt, S. Deser and C. W. Misner, {\it Gravitation:
An Introduction to Current Research}, edited by L. Witten, Wiley, New York
(1962).
\bibitem{d8} B. F. Schutz, Phys. Rev. D {\bf 2} (1970) 2762.
\bibitem{d9} B. F. Schutz, Phys. Rev. D {\bf 4} (1971) 3559.
\bibitem{bv} B. Vakili, Int.J.Mod.Phys.D {\bf 18} (2009) 1059.
\bibitem{bell} W. W. Bell, {\it Special Functions for Scientists and Engineers}, D.Van Nostrand Company Inc., London, (1968).

\bibitem{f11} A.B. Batista, J.C. Fabris, S.V.B. Gon\c{c}alves and J. Tossa, Phys. Rev. D {\bf 65} (2002) 063519;\\
 E. Farhi and S. Gutmann, Int. J. Mod. Phys. A {\bf 5} (1990) 3029.

\end{thebibliography}
\end{document}